\documentclass[aps,prc,preprint,tightenlines,superscriptaddress,showpacs,%
amssymb,byrevtex,nofootinbib]{revtex4}  %endfloats

\usepackage{epsfig,bm,dcolumn}

\begin{document}

\preprint{nucl-th/0306033}

%%%%%%%%%%%%%%%%%%%%% Title %%%%%%%%%%%%%%%%%%%%%%

\title{
$\bm{\rho}$ meson photoproduction at low energies}

%%%%%%%%%%%%%%%%%%%% Authors %%%%%%%%%%%%%%%%%%%%%
%%%%%%%%%%%%%%%%%%%% Addresses %%%%%%%%%%%%%%%%%%%%%

\author{Yongseok Oh}%
\email{yoh@phya.yonsei.ac.kr}

\affiliation{Institute of Physics and Applied Physics,
Yonsei University, Seoul 120-749, Korea}

\affiliation{Physics Division, Argonne National Laboratory, Argonne,
Illinois 60439}

\author{T.-S. H. Lee}%
\email{lee@phy.anl.gov}

\affiliation{Physics Division, Argonne National Laboratory, Argonne,
Illinois 60439}

\date{\today}

%%%%%%%%%%%%%%%%%%%% Abstract %%%%%%%%%%%%%%%%%%%%%

\begin{abstract}

The $\sigma$-exchange and $f_2$-exchange mechanisms for $\rho$ meson 
photoproduction are re-examined. 
Then the commonly employed $\sigma$-exchange amplitude is revised by using
the recent information from the analyses on the $\rho \to \pi^0\pi^0\gamma$
decay and the $\sigma NN$ coupling constant from Bonn potential.
Instead of relying on the Pomeron-$f$ proportionality assumption, the
$f_2$ meson exchange amplitude is established from an effective Lagrangian
which is constructed from the tensor structure of the $f_2$ meson.
Phenomenological information together with tensor meson dominance and
vector meson dominance assumptions are used to estimate the $f_2$ coupling
constants.
As a first step to improve the current theoretical models, we have also
explored the effects due to the un-correlated $2\pi$ exchange amplitude
with $\pi N$ intermediate state.
This leading-order $2\pi$ exchange amplitude can be calculated using the
coupling constants determined from the study of pion photoproduction and
the empirical width of $\rho \rightarrow \pi\pi$ decay.
In comparing with the existing differential cross section data, we find
that a model with the constructed $2\pi$, $\sigma$, and $f_2$ exchanges
is comparable to the commonly used $\sigma$ exchange model in which the
$\sigma$ coupling parameters are simply adjusted to fit the
experimental data.
We suggest that experimental verifications of the predicted single and
double spin asymmetries in the small $|t|$ ($ < 2$ GeV$^2$) region will be
useful for distinguishing the two models and improving our understanding of
the non-resonant amplitude of $\rho$ photoproduction.
Possible further improvements of the model are discussed.

\end{abstract}


\maketitle

\section{Introduction}

The recent experiments at Thomas Jefferson National Accelerator Facility
(TJNAF) \cite{CLAS00,CLAS01,CLAS01b,CLAS03}, GRAAL of Grenoble \cite{graal},
and LEPS of SPring-8 \cite{leps} are expected to provide new opportunities
for studying the electromagnetic production of vector mesons
at low energies.
For example, the differential cross section data for $\rho$
photoproduction from the CLAS Collaboration at TJNAF show big 
differences with the old data of 1970's \cite{CDGL76,BSYP78} in the large
momentum-transfer ($|t|$) region at low energies, where one may learn
about the $VNN$ couplings and other production mechanisms
\cite{TLTS99,Lage00,OB01,Oh03}.
Much more new data with similar high precisions will soon be available.

The study of vector meson photoproduction is expected to shed light
on the resolution of the so-called `missing resonance' problem
\cite{CR00,OTL01,ZLB98c,Zhao01,TL03}.
On the other hand, it is well known that this can be achieved only when
the nonresonant mechanisms are well understood \cite{OL02,PM02c}.
As a continuation of our effort in this direction \cite{OTL01,OL02}, we
explore in this work the nonresonant mechanisms of $\rho$ photoproduction.

There exist some investigations of the nonresonant mechanisms for vector
meson photoproduction.
To account for the diffractive features of the data in small $t$ region
at high energies, the Pomeron exchange model, as illustrated in
Fig.~\ref{fig:rho1}(a), was developed.
However, this model fails to describe the experimental observables at
low energies.
Indeed, meson exchanges (or secondary Reggeon exchanges) are found to be
crucial in understanding the low energy data.
In the case of $\omega$ photoproduction, it is well known that one-pion
exchange is the most dominant process at low energies. 
For $\rho$ photoproduction, however, the situation is not clear.
Generally, there are two scenarios which are based on either the $\sigma$
meson exchange model \cite{FS96,OTL00} or the $f_2$ exchange model
\cite{Lage00,KV01a-KV02}.
The $\sigma$ exchange model was motivated \cite{FS96} by the observation
that the decay width of $\rho\to\pi\pi\gamma$ is much larger than the
other radiative decays of the $\rho$ meson.
It is further assumed that the $\pi\pi$ in the $\pi\pi\gamma$ channel can
be modeled as a $\sigma$ meson so that the $\rho\sigma\gamma$ vertex can
be defined and modeled for calculating the $\sigma$ exchange mechanism as
illustrated in Fig.~\ref{fig:rho1}(b).
In practice, the product of the coupling constants 
$g_{\rho\sigma\gamma} g_{\sigma NN}$ of this tree-diagram is adjusted to
fit the cross section data of $\rho$ photoproduction at low energies.
If we use $g_{\sigma NN}^2/4\pi \sim 8$ from Bonn potential \cite{MHE87},
we then find that the resulting $g_{\rho\sigma\gamma}$ will yield a decay
width of $\rho^0 \to \sigma \gamma$ an order of magnitude larger than the
value extracted from the recent experimental decay width of
$\rho^0 \to \pi^0\pi^0 \gamma$ \cite{GY01a,BELN01,SND02}.
Thus the dynamical interpretation of the commonly used $\sigma$ meson
exchange model for $\rho$ meson photoproduction must be further examined
theoretically.

In this work, we would like to take a different approach to account for
the exchange of $\pi\pi$ in $\rho$ photoproduction.
First, the commonly employed $\sigma$-exchange amplitude is {\it revised\/}
by using the coupling constant $g_{\sigma NN}$ from Bonn potential and
$g_{\rho\sigma\gamma}$ from the recent experimental decay width of
$\rho^0 \to \pi^0\pi^0\gamma$ with the assumption \cite{FS96} that
$\pi^0\pi^0$ in this decay is strongly correlated and can be approximated
as a $\sigma$ particle. 
This is our starting point of developing a new model which is more
consistent with the existing meson-exchange models for $NN$ scattering
\cite{MHE87}, $\pi N$ scattering, and pion photoproduction \cite{SL96}.
We then consider the consequence of the strong $\rho \to \pi^+\pi^-$
decay which accounts for almost the entire decay width of the $\rho$ meson.
With the empirical value of the $\rho$ meson decay width, one can define
the $\rho\pi\pi$ vertex, which then leads naturally to the `uncorrelated'
two-pion exchange mechanism illustrated in Fig.~\ref{fig:2pi} with $M=\pi$
in the intermediate state.
A more complete calculation of un-correlated $2\pi$-exchange contributions
to $\rho$ photoproduction should also include other intermediate states
such as $\omega N$ and $\pi \Delta$.
However, the contributions from these intermediate states involve propagation
of two or three pions and must be considered along with other multi-pion
exchange mechanisms (such as the crossed diagrams due to the interchange of
$\gamma$ and $\rho$ lines in Fig.~\ref{fig:2pi}).
Obviously, this is a much more complex task and will not be attempted in
this exploratory investigation.
Our calculation of $2\pi$-exchange will be detailed in Section~II.F.

The $f_2$ exchange model for $\rho$ photoproduction was motivated
by the results from the analyses of $pp$ scattering data at low
energies \cite{DL92}.
In the study of $pp$ scattering the dominant secondary Regge trajectory
is represented by the $f$ trajectory, and the idea of Pomeron-$f$
proportionality had been used to model the Pomeron couplings using the $f_2$ 
couplings until 1970's \cite{Freu62,Freu71,CGZ71,KS73} before the advent
of the soft Pomeron model by Donnachie and Landshoff \cite{DL84}.
By considering the role of the $f$ trajectory in $pp$ scattering, it is
natural to consider the $f_2$ exchange model for vector meson photoproduction.
However, the $f_2$ exchange model developed in Refs. \cite{Lage00,KV01a-KV02}
for $\rho$ photoproduction made use of the Pomeron-$f$ proportionality in the
reverse direction. 
Namely, they assume that the structure of the $f_2$ couplings are the same
as that of the soft Pomeron exchange model.
Thus the $f_2$ tensor meson was treated as a $C=+1$ isoscalar photon, i.e.,
a vector particle.
In addition, the fit to the data is achieved by introducing an additional
adjustable parameter to control the strength of the $f_2$ coupling
\cite{Lage00}.
This is obviously not very satisfactory and leaves a room for
improvement.

Instead of relying on the Pomeron-$f$ proportionality assumption,
the $f_2$ meson exchange amplitude is evaluated in this work starting with
an effective Lagrangian which is constructed from the tensor structure of
the $f_2$ meson.
Phenomenological information together with tensor meson dominance and
vector meson dominance assumptions are used to estimate the $f_2$ coupling
constants.
With this, we then explore the extend to which the $\rho$ photoproduction
data can be described by a model that includes this newly constructed
$f_2$ exchange amplitude together with the revised $\sigma$-exchange
amplitude and the un-correlated $2\pi$-exchange amplitudes discussed above.

This paper is organized as follows.
In Section II, we explicitly define the amplitudes for the considered 
$\rho$ photoproduction mechanisms, including the Pomeron exchange,
$\sigma$ exchange, pseudoscalar meson exchanges, $s$- and $u$-channel
nucleon terms, and the newly constructed $f_2$ exchange.
The $2\pi$ exchange amplitudes are then given to complete our model 
construction.
The numerical results are presented in Section III.
For comparison, we consider two models.
Both models contain the $s$- and $u$-channel nucleon terms and
the exchanges of Pomeron, $\pi$, and $\eta$.
In addition, the first model includes the $\sigma$ exchange with free
parameters to fit the data following Refs. \cite{FS96,OTL00}, while
the second model contains the two-pion, $\sigma$, and $f_2$ exchanges,
where the parameters of the $\sigma$ exchange are fixed by Bonn
potential and $\rho\to \pi^0\pi^0\gamma$ decay.
We explore the extent to which these two rather different models can be
distinguished by examining the differential cross sections and spin
asymmetries.
Section IV contains a summary and discussions.
The details on the $f_2$ interactions with the photon and hadrons are
given in Appendix for completeness.

\section{Models for $\bm{\rho}$ photoproduction}

\begin{figure}
\centering
\epsfig{file=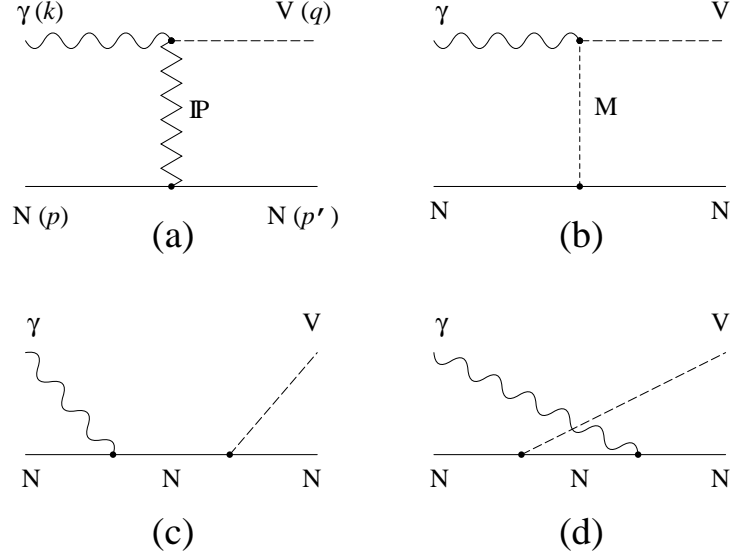, width=0.6\hsize}
\caption{Models for $\rho$ photoproduction. (a,b) $t$-channel Pomeron and
one-meson exchanges ($M=f_2,\pi,\eta,\sigma$), (c,d) $s$- and $u$-channel
nucleon pole terms.}
\label{fig:rho1}
\end{figure}

In this Section, we discuss possible production mechanisms for $\gamma p
\to \rho p$.
We first discuss single particle exchanges as depicted in Fig.~\ref{fig:rho1}.
Then the $2\pi$ exchange model will be constructed.
Each of the considered production amplitude, as illustrated in
Fig.~\ref{fig:rho1}, can be written as
\begin{equation}
T_{fi} = \varepsilon_\mu^*(V) \mathcal{M}^{\mu\nu}
\varepsilon_\nu^{}(\gamma),
\label{Tmat}
\end{equation}
where $\varepsilon_\mu^{}(V)$ and $\varepsilon_\nu^{}(\gamma)$ are the
polarization vectors of the vector meson and the photon, respectively.
We denote the four-momenta of the initial nucleon, final nucleon,
incoming photon, and outgoing vector meson by $p$, $p'$, $k$, and
$q$, respectively.
The Mandelstam variables are $s = W^2 = (k+p)^2$, $t = (p-p')^2$, and
$u = (p-q)^2$.

\subsection{Pomeron exchange}

We first consider the Pomeron exchange depicted in Fig.~\ref{fig:rho1}(a).
In this process, the incoming photon first converts into a $q\bar{q}$
pair, which interacts with the nucleon by the Pomeron exchange before
forming the outgoing vector meson.
The quark-Pomeron vertex is obtained by the Pomeron-photon analogy
\cite{DL84}, which treats the Pomeron as a $C=+1$ isoscalar photon, as
suggested by a study of nonperturbative two-gluon exchanges \cite{LN87}.
We then have \cite{DL84,LM95,PL97,TOYM98}
\begin{equation}
\mathcal{M}^{\mu\nu}_\mathbb{P} = G_\mathbb{P}^{}(s,t)
\mathcal{T}^{\mu\nu}_\mathbb{P},
\label{eq:MP}
\end{equation}
with
\begin{equation}
\mathcal{T}^{\mu\nu}_\mathbb{P} = i 12 \sqrt{4\pi\alpha_{\rm em}}
\frac{M_V^2 \beta_q \beta_{q'}}{f_V^{}} \frac{1}{M_V^2-t} \left(
\frac{2\mu_0^2}{2\mu_0^2 + M_V^2 - t} \right) F_1(t)
\bar{u}(p') \{ k\!\!\!/ \, g^{\mu\nu} - k^\mu \gamma^\nu \} u(p),
\label{M_P}
\end{equation}
where $\alpha_{\rm em} = e^2/4\pi$ and $F_1$ is the isoscalar
electromagnetic form factor of the nucleon,
\begin{equation}
F_1(t) = \frac{4M_N^2 - 2.8 t}{(4M_N^2 - t)(1-t/0.71)^2},
\end{equation}
with $t$ in GeV$^2$.
The proton and vector meson masses are represented by $M_N^{}$ and $M_V^{}$,
respectively. ($M_V^{} = M_\rho^{}$ in our case.)

The Regge propagator for the Pomeron in Eq. (\ref{eq:MP}) reads
\begin{equation}
G^{}_\mathbb{P} = \left(\frac{s}{s_0^{}}\right)^{\alpha^{}_P(t)-1}
\exp\left\{ - \frac{i\pi}{2} \left[ \alpha_P^{}(t)-1 \right]
\right\}.
\end{equation}
The Pomeron trajectory is taken to be the usual form
$\alpha_P^{} (t) = 1.08 + \alpha'_P t$ with $\alpha'_P = 1/s_0^{} = 0.25$
GeV$^{-2}$ \cite{DL84}.
In Eq.~(\ref{M_P}), $f_V$ is the vector meson decay constant:
$f_\rho = 5.33$, $f_\omega = 15.2$, and $f_\phi = 13.4$.
The coupling constants $\beta_u=\beta_d = 2.07$ GeV$^{-1}$,
$\beta_s = 1.60$ GeV$^{-1}$, and $\mu_0^2 = 1.1$ GeV$^2$ are chosen to
reproduce the total cross section data at high energies, $E_\gamma \ge 10$
GeV, where the total cross section of vector meson photoproductions are
completely dominated by the Pomeron exchange.
For $\rho$ photoproduction, we set
$\beta_q=\beta_{q'} = \beta_u = \beta_d$.

\subsection{$\bm{\sigma}$ meson exchange}

The $\sigma$ meson exchange model advocated by Friman and Soyeur
\cite{FS96} is based on the observation that $\Gamma(\rho\to \pi\pi\gamma)$
is the largest among all $\rho$ meson radiative decays, which
leads to the assumption that the $\rho$ photoproduction
process at low energies is dominated by the exchange of
$2\pi$. The $2\pi$ is then effectively represented by a $\sigma$ meson.
The effective Lagrangian for this model reads \cite{FS96,TLTS99,OTL00}
\begin{equation}
\mathcal{L}_\sigma = \frac{eg_{\rho\sigma\gamma}^{}}{M_\rho^{}} \left(
\partial^\mu \rho^\nu \partial_\mu A_\nu - \partial^\mu \rho^\nu \partial_\nu
A_\mu \right) \sigma + g_{\sigma NN}^{} \bar{N} \sigma N,
\label{eq:Lag-sigma}
\end{equation}
where $\rho_\mu$ is the $\rho^0$ meson field and $A_\mu$ the photon field.
The resulting $\sigma$ meson exchange amplitude is 
\begin{equation}
\mathcal{M}^{\mu\nu}_\sigma =
\frac{eg_{\rho\sigma\gamma} g_{\sigma NN}}{M_\rho^{}}
\frac{1}{t-M_\sigma^2} \left(k \cdot q g^{\mu\nu} - k^\mu q^\nu \right)
\bar{u}(p') u(p) F_{\sigma NN}(t) F_{\rho\sigma\gamma}(t),
\end{equation}
where  
\begin{equation}
F_{\sigma NN}(t) = \frac{\Lambda_\sigma^2 - M_\sigma^2}{\Lambda_\sigma^2
- t}, \qquad
F_{\rho\sigma\gamma}(t) = \frac{\Lambda_{\rho\sigma\gamma}^2 - M_\sigma^2}
{\Lambda_{\rho\sigma\gamma}^2 - t}
\end{equation}
are the form factors.
The cutoff parameters of the form factors and the product of coupling
constants $g_{\rho\sigma\gamma} g_{\sigma NN}$ are adjusted to fit the
$\rho$ photoproduction data at low energies.
It was found \cite{FS96,OTL00} that
\begin{eqnarray}
&& M_\sigma = 0.5 \mbox{ GeV}, \qquad g^2_{\sigma NN}/4\pi = 8.0, \qquad
g_{\rho\sigma\gamma}^{} = 3.0, \nonumber \\
&& \Lambda_\sigma = 1.0 \mbox{ GeV}, \qquad \Lambda_{\rho\sigma\gamma} = 0.9
\mbox{ GeV}.
\label{para:sigma}
\end{eqnarray}
The resulting $\sigma$ mass parameter is close to the value
$M_\sigma= 0.55 \sim 0.66$ GeV of Bonn potential \cite{MHE87}.
If we further take the value $g^2_{\sigma NN}/4\pi = 8.3 \sim 10$ from Bonn
potential, we then find that the resulting $g_{\rho\sigma\gamma}$ is close
to the values from the QCD sum rules,
$ g_{\rho\sigma\gamma}^{}(\mbox{\small QCDSR}) = 3.2 \pm 0.6$
\cite{GY01} or $2.2 \pm 0.4$ \cite{AOS02}.
However such a large value of $g_{\rho\sigma\gamma}^{}$ corresponds to 
the $\rho \rightarrow \sigma \gamma (\to \pi^0 \pi^0 \gamma)$ decay width
that is much larger than the empirical value of
$\Gamma(\rho^0 \to \pi^0 \pi^0 \gamma)$ \cite{BELN01,SND02}.
If we accept the empirically estimated, but model-dependent value of SND
experiment \cite{SND02}, $\mbox{BR}(\rho \to \sigma
\gamma) = (1.9 \stackrel{+ 0.9}{\mbox{\scriptsize $-0.8$}} \pm 0.4)
\times 10^{-5}$, which gives $\Gamma(\rho \to \sigma \gamma) \approx
2.83$ keV, we get
\begin{equation}
|g_{\rho\sigma\gamma}| \approx 0.25,
\end{equation}
since the Lagrangian (\ref{eq:Lag-sigma}) gives
\begin{equation}
\Gamma(\rho \to \sigma\gamma) = \frac{\alpha_{\rm em}^{}
g^2_{\rho\sigma\gamma}}{24 M_\rho^5} \left( M_\rho^2 - M_\sigma^2
\right)^3.
\end{equation}
This value is smaller than that of Eq.~(\ref{para:sigma}) by an order
of magnitude.
Therefore, the $\sigma$ exchange model suffers from the big uncertainty
of $g_{\rho\sigma\gamma}$, which is under debate
\cite{BELN01,GY01a,SND02,BE03,OK03}.
Furthermore, there is no clear particle identification of a $\sigma$
particle and the use of $\sigma$ exchange in defining $NN$ potential
has been seriously questioned.
Thus it is possible that the $\sigma$ exchange may not be the right major
mechanism for $\rho$ photoproduction.

\subsection{Pseudoscalar meson exchanges}

The $\pi$ and $\eta$ meson exchanges are also allowed for $\rho$
photoproduction, although their contributions are known to be not important.
They are calculated from
\begin{eqnarray}
\mathcal{L}_{\rho\gamma\varphi} &=& \frac{eg_{\rho\gamma\varphi}}{M_V}
\epsilon^{\mu\nu\alpha\beta} \partial_\mu \rho_\nu \partial_\alpha
A_\beta \varphi,
\nonumber \\
\mathcal{L}_{\varphi NN} &=& \frac{g_{\varphi NN}}{2M_N} \bar{N} \gamma^\mu
\gamma_5 \partial_\mu \varphi N,
\label{lag:pi}
\end{eqnarray}
where $\varphi = \pi^0, \eta$.
The coupling constants $g_{\rho\gamma\varphi}$ are fixed by the $\rho \to
\varphi \gamma$ decay widths
\begin{equation}
\Gamma(\rho \to \varphi\gamma) = \frac{\alpha_{\rm em}
g_{\rho\gamma\varphi}^2}{24 M_V^5} (M_V^2 - M_\varphi^2)^3.
\end{equation}
Using the experimental data \cite{PDG02},
$\Gamma(\rho^0 \to \pi^0 \gamma)_{\rm expt.} = 121 \pm 31$ keV and
$\Gamma(\rho^0 \to \eta \gamma)_{\rm expt.} = 62 \pm 17 $ keV,
we get
\begin{equation}
g_{\rho\gamma\pi} = 0.756, \qquad g_{\rho\gamma\eta} = 1.476.
\end{equation}
This also gives $g_{\omega\gamma\pi} = 1.843$ and
$g_{\omega\gamma\eta} = 0.414$.
We use $g_{\pi NN}^2/4\pi = 14.3$ and the SU(3) relation to get
$g_{\eta NN}^2/4\pi = 0.99$.
Although there are other estimates on the value of $g_{\eta NN}$ reported
in the literature, the role of the $\eta$ exchange is much suppressed in
$\rho$ photoproduction and the dependence of our results on $g_{\eta
NN}$ is negligible.

The pseudoscalar meson exchange amplitude, Fig.~\ref{fig:rho1}(b),
calculated from the Lagrangian (\ref{lag:pi}) reads
\begin{equation}
\mathcal{M}_\varphi^{\mu\nu} =
\frac{i e g_{\rho\gamma\varphi} g_{\varphi NN}}{2M_N M_V}
\frac{1}{t-M_\varphi^2} \epsilon^{\mu\nu\alpha\beta} q_\alpha k_\beta
\bar{u}(p') (p\!\!\!/ - p\!\!\!/') \gamma_5 u(p) F_{\varphi NN}(t)
F_{\rho\varphi\gamma}(t),
\end{equation}
where the form factors are
\begin{equation}
F_{\varphi NN}(t) = \frac{\Lambda_\varphi^2 -
M_\varphi^2}{\Lambda_\varphi^2 - t}, \qquad
F_{\rho\varphi\gamma}(t) =
\frac{\Lambda_{\rho\varphi\gamma}^2 - M_\varphi^2}
{\Lambda_{\rho\varphi\gamma}^2 - t}.
\end{equation}
We use $\Lambda_\pi = 0.6$ GeV, $\Lambda_{\rho\pi\gamma} = 0.77$ GeV,
$\Lambda_\eta = 1.0$ GeV, and $\Lambda_{\rho\eta\gamma} = 0.9$ GeV
\cite{FS96,OTL01}.

\subsection{Nucleon pole terms}

The $s$- and $u$-channel nucleon terms, Figs.~\ref{fig:rho1}(c,d), are
calculated from
\begin{eqnarray}
\mathcal{L}_{\gamma pp} &=& - e \bar{N} \left[ A_\mu \gamma^\mu -
\frac{\kappa_p}{2M_N^{}} \sigma_{\mu\nu} \partial^\nu A^\mu \right] N,
\nonumber \\
\mathcal{L}_{\rho pp} &=& - \frac{g_{\rho NN}}{2}
\bar{N} \left[ \rho^\mu \gamma_\mu
- \frac{\kappa_\rho}{2M_N^{}} \sigma_{\mu\nu} \partial^\nu \rho^\mu
\right] N.
\end{eqnarray}
The resulting production amplitude is
\begin{eqnarray}
\mathcal{M}_N^{\mu\nu} &=& \frac{eg_{\rho NN}}{2} \bar{u}(p')
\biggl[ \Gamma^\mu_V(q) \frac{p\!\!\!/
+ k\!\!\!/ + M_N}{s-M_N^2} \Gamma_\gamma^\nu(k) F_N(s)
\nonumber \\ && \qquad \mbox{}
 + \Gamma^\nu_\gamma(k) \frac{p\!\!\!/
- q\!\!\!/ + M_N}{u-M_N^2} \Gamma_V^\mu(q) F_N(u) \biggr] u(p),
\end{eqnarray}
where
\begin{equation}
\Gamma^\mu_V = \gamma^\mu - i \frac{\kappa_\rho}{2M_N} \sigma^{\mu\nu}
q_\nu, \qquad
\Gamma^\mu_\gamma = \gamma^\mu + i \frac{\kappa_p}{2M_N} \sigma^{\mu\nu}
k_\nu.
\end{equation}
The form factor has the form \cite{PJ91}
\begin{equation}
F_N(r) = \frac{\Lambda_N^4}{\Lambda_N^4 + (r-M_N^2)^2},
\end{equation}
with $\Lambda_N = 0.5$ GeV
taken from Refs. \cite{OTL01,TLTS99}. This choice of the
nucleon form factor leads to a satisfactory explanation of
the steep rise of the differential
cross sections with increasing $|t|$ in terms of the $u$-channel nucleon 
term [Fig.~\ref{fig:rho1}(d)].

Because $F_N(s) \neq F_N(u)$, the above amplitude does not satisfy the
gauge invariance. In order to restore the gauge invariance, we project
out the gauge non-invariant terms as
\begin{equation}
\Gamma_V^\mu \to \Gamma_V^\mu - \frac{k^\mu}{k \cdot q} \,
q \cdot \Gamma_V,
\qquad
\Gamma_\gamma^\mu \to \Gamma_\gamma^\mu - \frac{q^\mu}{k \cdot q} \,
k \cdot \Gamma_\gamma.
\end{equation}
For the $\rho NN$ coupling constants, we take the values determined in
the analyses of pion photoproduction and $\pi N$ scattering \cite{SL96}:
\begin{equation}
g_{\rho NN} = 6.2, \qquad \kappa_\rho = 1.0,
\end{equation}
and the anomalous magnetic moment of the nucleon is $\kappa_p = 1.79$.

\subsection{$\bm{f_2}$ meson exchange}

We now discuss the exchange of the $f_2(1270)$ tensor meson, which has
quantum numbers $I^G(J^{PC}) = 0^+(2^{++})$.
The mass and decay width of the $f_2(1270)$ are $M_f = 1275.4 \pm 1.2$
MeV and $\Gamma(f_2) = 185.1 \stackrel{+ 3.4}{\mbox{\scriptsize $-2.6$}}$
MeV \cite{PDG02}.
Because of its quantum numbers, it has been once suggested as a candidate
for the Pomeron.
But this assumption violates the duality with the $a_2$ trajectory which
includes $I^G(J^{PC})=1^-(2^{++})$ state and it is now believed that the
$f_2$ does not lie on the Pomeron trajectory.

In the approach of Ref.~\cite{Lage00}, the $f_2$ is treated as a $C=+1$
isoscalar photon just like the Pomeron. This leads to
a Regge amplitude of the following form
\begin{equation}
\mathcal{M}^{\mu\nu}_{f_2^{}} = \kappa_{f_2^{}}^{} G_{f_2^{}}^{}(s,t)
\mathcal{T}^{\mu\nu}_\mathbb{P},
\end{equation}
where%
\footnote{
The form of $G_{f_2^{}}$ in Eq.~(\ref{f2-tr}) is due to the fact that
the $f_2$ interaction is treated as that of an isoscalar photon, i.e.,
a vector particle interaction. If we use the tensor structure of the $f_2$
interaction, it would be 
$$ G_{f_2}^{}(s,t) = \left( \frac{s}{s_1^{}} \right)^{\alpha_{f_2}(t) - 2}
\frac{(1+\exp[-i\pi\alpha_{f_2}^{}(t)])\pi \alpha'_{f_2}}{ 2 \sin[\pi
\alpha_{f_2}^{}(t)] \, \Gamma[ \alpha_{f_2}^{}(t)-1]}. $$}
\begin{equation}
G_{f_2^{}}^{}(s,t) = \left( \frac{s}{s_1^{}} \right)^{\alpha_{f_2}(t) - 1}
\frac{(1+\exp[-i\pi\alpha_{f_2}^{}(t)])\pi \alpha'_{f_2}}{ 2 \sin[\pi
\alpha_{f_2}^{}(t)] \, \Gamma[ \alpha_{f_2}^{}(t)]},
\label{f2-tr}
\end{equation}
with $s_1^{} = 1/\alpha'_{f_2} \approx 1$ GeV$^2$, while the form of
$\mathcal{T}^{\mu\nu}_\mathbb{P}$ is the same as given in Eq.~(\ref{M_P}).
The $f_2$ trajectory is linearly approximated as $\alpha_{f_2}(t)
\approx 0.47 + 0.89 t$ \cite{DL92,KV01a-KV02}.
In order to control the strength of the $f_2$ couplings to the hadrons, a
free parameter $\kappa_{f_2^{}}^{}$ was introduced \cite{Lage00} and
adjusted to fit the $\rho$ photoproduction data at low energies.

In this paper, we depart from this Regge parameterization and construct an
$f_2$ exchange model solely based on the tensor structure of the $f_2$ meson.
We will use the experimental data associated with the $f_2$ meson, the tensor
meson dominance, and vector meson dominance assumptions to fix the $f_2$
coupling constants, such that the strength of the resulting $f_2$ exchange
amplitude is completely fixed in this investigation.
Following Refs. \cite{Gold68,PSMM73}, the effective Lagrangian accounting
for the tensor structure of the $f_2 NN$ interaction is written as%
\footnote{In the conventions of Ref.~\cite{PSMM73}, $G_{fNN}^{(1)}=G_{fNN}^{}$
and $G_{fNN}^{(2)} = F_{fNN}^{}$.}
\begin{equation}
\mathcal{L}_{fNN} = -2i \frac{G_{fNN}^{}}{M_N} \bar{N} (\gamma_\mu
\partial_\nu + \gamma_\nu \partial_\mu ) N f^{\mu\nu}
+ 4 \frac{F_{fNN}^{}}{M_N} \partial_\mu \bar{N} \partial_\nu N
f^{\mu\nu},
\end{equation}
where $f^{\mu\nu}$ is the $f_2$ meson field.
This gives the following form of the $fNN$ vertex function,
\begin{equation}
V_{be,d} = -\epsilon^{\mu\nu} \bar{u}(p_d) \left\{ \frac{G_{fNN}^{}}{M_N}
\left[ \Sigma_\mu \gamma_\nu + \gamma_\nu \Sigma_\mu \right] +
\frac{F_{fNN}^{}}{M_N^2} \Sigma_\mu \Sigma_\nu \right\} u (p_b),
\label{VfNN}
\end{equation}
where $\Sigma_\mu = (p_b+p_d)_\mu$, $p_b$ and $p_d$ are the incoming
and outgoing nucleon momentum, respectively, and $\epsilon^{\mu\nu}$ is the
polarization tensor of the $f_2$ meson.

The coupling constants associated with the $f_2$ meson were first estimated
by using the dispersion relations to analyze the backward $\pi N$ scattering
\cite{Gold68} and the $\pi\pi\to N\bar{N}$ partial-wave amplitudes.
The results are summarized in Table \ref{tab:fnn}.
Note that the value estimated based on the tensor-meson dominance
\cite{Renn70} is much smaller than the empirical values.
(See Appendix for details.)

\begin{table}
\centering
\begin{tabular}{c|c|c} \hline\hline
${G_{fNN}^2}/{4\pi}$ & $F_{fNN}^{}/G_{fNN}^{}$ &
     \\ \hline
$1.12$  &  ---  & Ref. \cite{Gold68} \\
$3.31$ & $\approx 0$ & Ref. \cite{ASS70} \\
$3.31 \pm 0.63$ & $0.06 \pm 0.17$ & Ref. \cite{Engels70} \\
$4.0 \pm 1.0$ & $0.00 \pm 0.07$ & Ref. \cite{Hede77} \\
$2.2 \pm 0.9$ & $0.6 \pm 0.9$ & Ref. \cite{BK77} \\ \hline
$0.38 \pm 0.04$ & $\approx 0$ & Ref. \cite{Renn70} \\
\hline\hline
\end{tabular}
\caption{Estimates on the $fNN$ coupling constants, $G_{fNN}$ and
$F_{fNN}$, using $\pi N$ dispersion relations. The values are compared
with the prediction of tensor meson dominance \cite{Renn70}.}
\label{tab:fnn}
\end{table}

The most general form for the $fV\gamma$ vertex satisfying gauge
invariance reads \cite{Renn71}
\begin{equation}
\langle \gamma(k) V(k') | f_2 \rangle = \frac{1}{M_f}
\epsilon^\kappa \epsilon'^\lambda f^{\mu\nu} A_{\kappa \lambda
\mu\nu}^{fV\gamma}(k,k'),
\end{equation}
where $\epsilon$ and $\epsilon'$ are the polarization vectors of the
photon and the vector meson, respectively, and
\begin{eqnarray}
A^{fV\gamma}_{\kappa\lambda\mu\nu} (k,k') &=&
\frac{f^{}_{fV\gamma}}{M_f^3} \left[ g_{\kappa \lambda} (k\cdot k')
- k'_\kappa k_\lambda \right] (k-k')_\mu (k-k')_\nu
\nonumber \\ && \mbox{}
+ g_{fV\gamma}^{} [ g_{\kappa\lambda} (k-k')_\mu (k-k')_\nu
+ g_{\lambda\mu} k'_\kappa (k-k')_\nu
+ g_{\lambda\nu} k'_\kappa (k-k')_\mu
\nonumber \\ && \mbox{} \qquad
- g_{\kappa\mu} k_\lambda (k-k')_\nu
- g_{\kappa\nu} k_\lambda (k-k')_\mu
-2 k \cdot k' ( g_{\kappa\mu} g_{\lambda\nu} + g_{\kappa\nu}
g_{\lambda\mu}) ].
\label{AFVgamma}
\end{eqnarray}
The tensor meson dominance assumption together with the vector meson
dominance gives
\cite{Renn71}
\begin{equation}
f^{}_{fV\gamma} = 0 \qquad \mbox{ and } \qquad
g^{}_{fV\gamma} = \frac{e}{f_V} G_{fVV},
\label{fvg:coup}
\end{equation}
where
\begin{equation}
G_{fVV} = G_{f\pi\pi} = 5.76.
\end{equation}
Here $G_{f\pi\pi}$ is determined from the decay width of $f_2 \to
\pi\pi$.
The details on the $f_2$ interactions with the photon and hadrons, and
tensor meson dominance are given in Appendix.

With the above formulas, it is straightforward to obtain the production
amplitude as
\begin{equation}
\mathcal{M}^{\mu\nu}_{f_2} = - \bar{u}(p) \Gamma^{\alpha\beta}(p,p') u(p)
\frac{P_{\alpha\beta;\rho\sigma}}{(p-p')^2 - M_f^2}
V^{\rho\sigma;\nu\mu}(k,q) F_{fNN}(t) F_{fV\gamma}(t),
\label{eq:f2-amp}
\end{equation}
where
\begin{eqnarray}
\Gamma_{\alpha\beta}(p,p') &=& \frac{G_{fNN}}{M_N} \left[ (p+p')_\alpha
\gamma_\beta + (p+p')_\beta \gamma_\alpha \right]
+ \frac{F_{fNN}}{M_N^2} (p+p')_\alpha (p+p')_\beta,
\nonumber \\
P_{\alpha\beta;\rho\sigma} &=& \frac12 (\bar{g}_{\alpha\rho}
\bar{g}_{\beta\sigma} + \bar{g}_{\alpha\sigma} \bar{g}_{\beta\rho}) -
\frac13 \bar{g}_{\alpha\beta} \bar{g}_{\rho\sigma},
\nonumber \\
V^{\rho\sigma;\nu\mu}(k,q) &=& \frac{f_{fV\gamma}^{}}{M_f^4} \left[ -
g_{\mu\nu} (k \cdot q) + q_\nu k_\mu \right] (k+q)_\rho (k+q)_\sigma
\nonumber \\ && \mbox{}
+ \frac{g_{fV\gamma}^{}}{M_f} \biggl[ g_{\mu\nu} (k+q)_\rho (k+q)_\sigma
- g_{\mu\rho} q_\nu (k+q)_\sigma - g_{\mu\sigma} q_\nu (k+q)_\rho
\nonumber \\ && \mbox{} \qquad
- g_{\nu\rho} k_\mu (k+q)_\sigma - g_{\nu\sigma} k_\mu (k+q)_\rho
+ 2 k \cdot q ( g_{\nu\rho} g_{\mu\sigma} + g_{\nu\sigma} g_{\mu\rho})
\biggr],
\end{eqnarray}
and
\begin{equation}
\bar{g}_{\mu\nu} = - g_{\mu\nu} + \frac{(p-p')_\mu (p-p')_\nu}{M_f^2}.
\end{equation}
The form factors are chosen as 
\begin{equation}
F_{f NN}(t) = \frac{\Lambda_{fNN}^2 - M_{f_2}^2}{\Lambda_{fNN}^2
- t}, \qquad
F_{fV\gamma}(t) = \frac{\Lambda_{fV\gamma}^2 - M_{f_2}^2}
{\Lambda_{fV\gamma}^2 - t},
\end{equation}
where the cutoff parameters will be discussed in Sect. III.
The relative phases among $f_2$ couplings are fixed by tensor meson
dominance.

\subsection{$\bm{2\pi}$ exchange}

\begin{figure}
\centering
\epsfig{file=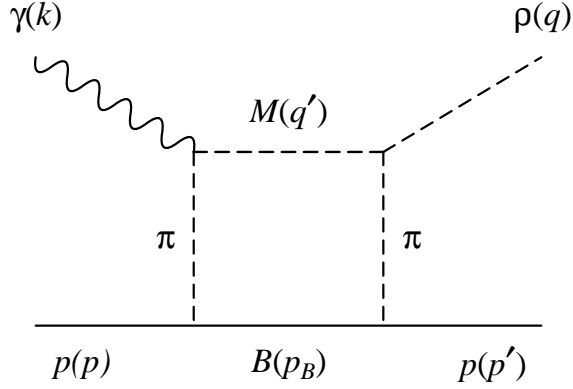, width=0.5\hsize}
\caption{$2\pi$ exchange in $\rho$ photoproduction. The intermediate meson
state ($M$) includes $\pi$, and the baryon ($B$) includes the nucleon.}
\label{fig:2pi}
\end{figure}

In this subsection, we discuss the $2\pi$ exchange for $\rho$
photoproduction as shown in Fig.~\ref{fig:2pi}.
We only consider the contribution from setting the intermediate state
$(MB) = (\pi N)$.
As discussed in Section I, the contributions from other intermediate
states like $\pi\Delta$ and $\omega N$ involve propagation of two or three
pions and hence are neglected along with the other multi-meson exchange
amplitudes in this exploratory investigation.

We compute the loop amplitude of Fig.~\ref{fig:2pi} by making use of
the method of Sato and Lee \cite{SL96}, which gives
\begin{equation}
T_{\rm loop} = \int d^3 {\bf q}' \left[ \varepsilon(\gamma) \cdot
B_{\gamma N, M N} ({\bf k}, {\bf q}';E) \right] G_{M N}({\bf q}',E)
\left[ \varepsilon^*(V) \cdot V_{M N,\rho N} ({\bf q}',{\bf q}:E)
\right],
\label{eq:loop}
\end{equation}
where
\begin{equation}
G_{M N}({\bf q}',E) = \frac{1}{E-E_N(q')-E_M (q')+i\epsilon}.
\label{Green}
\end{equation}
Obviously, $B_{\gamma N, M N}$ and $V_{M N, \rho N}$ are the
one-pion-exchange amplitudes illustrated in Fig.~\ref{fig:2pi}.
We only consider $(MN)=(\pi N)$ intermediate state in this paper.

Equation~(\ref{eq:loop}) can be rewritten as
\begin{eqnarray}
T_{\rm loop} &=& \mathcal{P} \int d^3 {\bf q}' \frac{\varepsilon^*(V)
\cdot V({\bf q}', {\bf q})\, \varepsilon(\gamma) \cdot B({\bf k},{\bf q}')}
{W - E_B(q') - E_M(q')}
\nonumber \\ && \mbox{} - i \int d\Omega_{k_t} \,\rho_{BM}^{}({\bf
k}_t)\,
\varepsilon^*(V) \cdot V({\bf k}_t,{\bf q}) \, \varepsilon(\gamma) \cdot
B({\bf k},{\bf k}_t) \, \theta(W-M_M-M_B),
\end{eqnarray}
where the subscripts of $V$ and $B$ are understood.
Here $\theta(x)$ is the step function and
\begin{equation}
\rho_{BM}^{}(k) = \frac{\pi k E_B(k) E_M(k)}{E_B(k) + E_M(k)},
\end{equation}
where $E_B(k)$ and $E_M(k)$ are the energies of the intermediate
baryon and meson with momentum ${\bf k}$.
Through the on-shell condition $W = E_B(k_t) + E_M(k_t)$, $k_t$ is
determined as
\begin{equation}
k_t = \frac{1}{2W} \sqrt{ \lambda(W^2,M_M^2,M_B^2)},
\end{equation}
where
\begin{equation}
\lambda(x,y,z) = x^2 + y^2 + z^2 - 2(xy+yz+zx).
\end{equation}

For the considered $(MN)=(\pi N)$ case, 
the one-pion-exchange amplitudes $B_{\gamma N, \pi N}$
and $V_{\pi N,\rho N}$ in Eq.(35) 
can be calculated from
\begin{eqnarray}
\mathcal{L}_{\gamma\pi\pi} &=& e \left[ \partial^\mu \bm{\pi} \times
\bm{\pi} \right]_3 A_\mu,
\nonumber \\
\mathcal{L}_{\rho\pi\pi} &=& g_{\rho\pi\pi} \bm{\rho}_\mu \cdot \left(
\bm{\pi} \times \partial^\mu \bm{\pi} \right),
\nonumber \\
\mathcal{L}_{\pi NN} &=& \frac{g_{\pi NN}}{2M_N} \bar{N} \gamma^\mu
\gamma_5 \bm{\tau} \cdot \partial_\mu \bm{\pi} N.
\label{eq:lags}
\end{eqnarray}
The coupling constant $g_{\rho\pi\pi}$ is determined from the decay
width $\Gamma(\rho\to \pi\pi)$, which reads
\begin{equation}
\Gamma(\rho\to\pi\pi) =
\frac{g_{\rho\pi\pi}^2}{48 \pi M_\rho^2} \left(
M_\rho^2 - 4 M_\pi^2 \right)^{3/2}.
\end{equation}
Using $\Gamma(\rho^0 \to \pi^+\pi^-) = 150.7$ MeV \cite{PDG02}, we
obtain
\begin{equation}
g_{\rho\pi\pi} = 6.04.
\end{equation}
Then the $2\pi$-exchange transition amplitude with intermediate $\pi N$
channel reads
\begin{eqnarray}
\widetilde{\mathcal{M}}_{\pi N}^{\mu\nu} & \equiv & V^\mu(q',q) B^\nu(k,q')
\nonumber \\  &=& 
\frac{1}{(2\pi)^3} \frac{M_N}{E_N(p_B)}
\frac{1}{2E_\pi(q')} \frac{e g_{\rho\pi\pi} g_{\pi NN}^2}{4M_N^3}
(q'-p_B+p')^\mu (p_B - p -q')^\nu
\nonumber \\ && \mbox{} \times
\frac{1}{(p_B-p)^2-M_\pi^2} \frac{1}{(p_B-p')^2 - M_\pi^2} \bar{u}(p')
\Gamma u(p),
\label{piN}
\end{eqnarray}
where
\begin{equation}
\Gamma = (p\!\!\!/' - p_B^{}\!\!\!\!\!\!/\ \ ) \left( p_B^{}\!\!\!\!\!\!/\ \
- M_N \right) (p_B^{}\!\!\!\!\!\!/\ \  - p\!\!\!/).
\label{Gamma}
\end{equation}
The loop integration must be regularized by introducing form factors.
We include the form factors for each vertices.
In addition, we also introduce the form factor to take into account the
off-shell-ness of the intermediate states,
\begin{equation}
F_\ell ({\bf q}') = \left( \frac{\Lambda_\ell^2 + {\bf k}_t^2}{\Lambda_\ell^2
+ {\bf q}'^2} \right)^2.
\label{eq:Fell}
\end{equation}
Thus the final form of the form factor is
\begin{equation}
F = F_\ell ({\bf q}') F_{\rho\pi\pi} (t_1) F_{\rho\pi\pi}(t_2) F_{\pi NN}
(t_1) F_{\pi NN}(t_2),
\label{eq:FF}
\end{equation}
where
\begin{equation}
F_{\rho\pi\pi}(t) = \frac{\Lambda_{\rho\pi\pi}^2 -
M_\pi^2}{\Lambda_{\rho\pi\pi}^2 - t}, \qquad
F_{\pi NN}(t) = \frac{\Lambda_{\pi NN}^2 -
M_\pi^2}{\Lambda_{\pi NN}^2 - t},
\end{equation}
and $t_1 = (p_B^{}-p)^2$ and $t_2 = (p_B^{}-p')^2$.
Here the inclusion of $F_{\rho\pi\pi}(t_1)$ implies the vector meson
dominance assumption.
The cutoff parameters will be discussed in Sect. III.

We now comment on the loop calculation described above.
We do not consider the crossed diagrams of Fig.~\ref{fig:2pi}, since such
diagrams include three-particle intermediate states and hence are of
higher-order effects which are neglected in this exploratory study.
However by neglecting the crossed diagrams, the resulting amplitude does
not satisfy gauge invariance.
In this study, therefore, we restore gauge invariance of the amplitude
(\ref{piN}) by projecting out the gauge non-invariant terms as \cite{TOYM98}
\begin{equation}
\widetilde{\mathcal{M}}^{\mu \nu} \to \mathcal{P}^{\mu\mu'}
\widetilde{\mathcal{M}}_{\mu'\nu'} \mathcal{P}^{\nu'\nu},
\end{equation}
where the projection operator reads
\begin{equation}
\mathcal{P}^{\mu\nu} = g^{\mu\nu} - \frac{k^\mu q^\nu}{k \cdot q}.
\end{equation}

\section{Cross sections and polarization asymmetries}

In this work we first re-examine the commonly employed $\sigma$ exchange
by considering model (A) which includes the Pomeron, $\sigma$, $\pi$, $\eta$
exchanges, and the $s$- and $u$-channel nucleon terms.
We then explore model (B) which is constructed by replacing the $\sigma$
exchange in model (A) by the $f_2$ and $2\pi$ exchanges.
We also add the $\sigma$ exchange to model (B) as a correlated $2\pi$
exchange with the couplings determined by $\rho\to\pi^0\pi^0\gamma$ decay
and Bonn potential.
All parameters of the models are explained in Section II.
In particular, the $\sigma$-exchange parameters in model (A) are given in
Eq.~(\ref{para:sigma}), which are chosen to reproduce the $\rho$
photoproduction data.

For model (B), we use the $f_2$ couplings as (see Appendix)
\begin{equation}
{G_{fNN}^2}/{4\pi} = 2.2, \qquad F_{fNN} = 0, \qquad G_{fVV} = 5.76,
\label{eq:cut}
\end{equation}
with the relation (\ref{fvg:coup}).
The recently estimated $\Gamma(\rho\to\sigma\gamma)$ \cite{SND02} is
used to constrain $g_{\rho\sigma\gamma}$ as
\begin{equation}
g_{\rho\sigma\gamma} = 0.25.
\end{equation}
The other parameters for the $\sigma$ exchange are the same as given in
Eq.~(\ref{para:sigma}).
The only unspecified parameters are the cutoff parameters
$\Lambda_{fNN}$ and $\Lambda_{fV\gamma}$ for the $f_2$ exchange and
the cutoff parameters of Eq.~(\ref{eq:FF}) for regularizing the loop
integrations.   
The parameter $\Lambda_\ell$ for all loop integrations is fixed to be 
$0.5$ GeV which is identical to the value used in our previous 
investigation \cite{OL02} of the one-loop corrections on $\omega$
photoproduction.
The other cutoffs including $\Lambda_{\rho\pi\pi}$ and $\Lambda_{\pi NN}$
in the loop calculation are chosen to be 0.6 GeV.
The other two parameters of model (B) are adjusted to fit the cross
section data and are found to be 
\begin{eqnarray}
\Lambda_{fNN} &=& \Lambda_{fV\gamma} = 1.4 \mbox{ GeV}.
\end{eqnarray}
This is a unsatisfactory aspect of this work, but it is unavoidable in any
phenomenological approach.
Future theoretical calculations of form factors are therefore highly desirable.

\begin{figure}
\centering
\epsfig{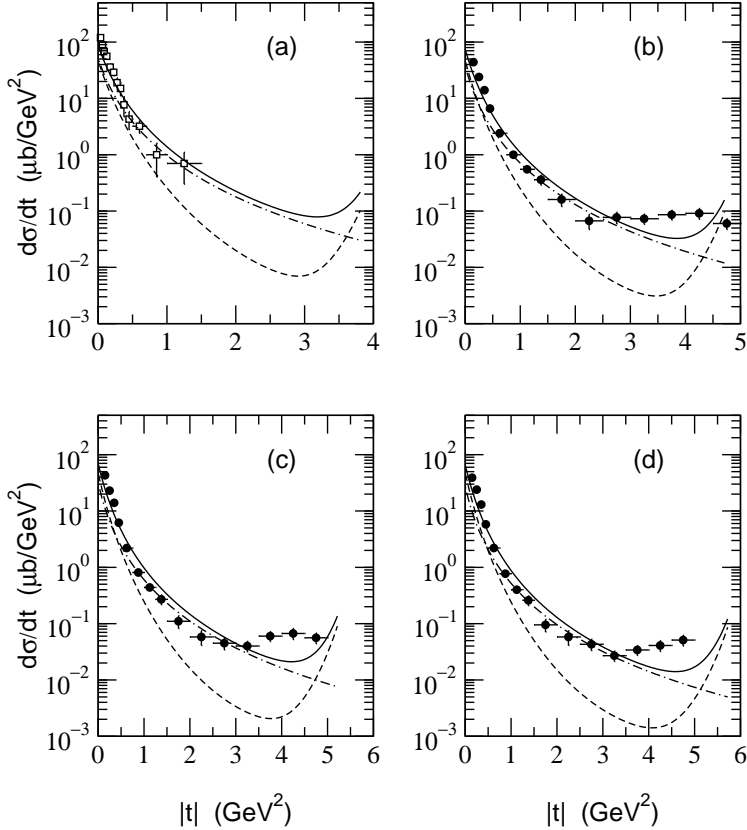}
\caption{Differential cross sections of model (A) at $E_\gamma = $ (a) 2.8,
(b) 3.28, (c) 3.55, and (d) 3.82 GeV. The dot-dashed lines are from $\sigma$
exchange and the dashed lines are without $\sigma$ exchange. The solid
lines are the full calculation. Experimental data are from
Ref.~\cite{BCGG72} (open squares) and Ref.~\cite{CLAS01b} (filled circles).}
\label{fig:model-A}
\end{figure}

The differential cross sections for $\gamma p \to \rho p$ calculated from
model (A) are compared with the SLAC data \cite{BCGG72} and the recent CLAS
data \cite{CLAS01b} in Fig.~\ref{fig:model-A}.
We see that the full calculations (solid curves) are dominated by the
$\sigma$ exchange contributions (dot-dashed curves).
The contributions from the other exchange mechanisms (dashed curves)
become comparable only in the very forward and backward angles.
This is mainly due to the fact that the Pomeron exchange
[Fig.~\ref{fig:rho1}(a)] is forward peaked and the $u$-channel nucleon term
[Fig.~\ref{fig:rho1}(d)] is backward peaked.
It is clear that the data can only be qualitatively reproduced by model (A).
The main difficulty is in reproducing the data in the large $|t|$ (larger than
about 3 GeV$^2$) region.
No improvement can be found by varying the cutoff parameters of various form
factors of model (A).
This implies the role of other production mechanisms in this region.

\begin{figure}
\centering
\epsfig{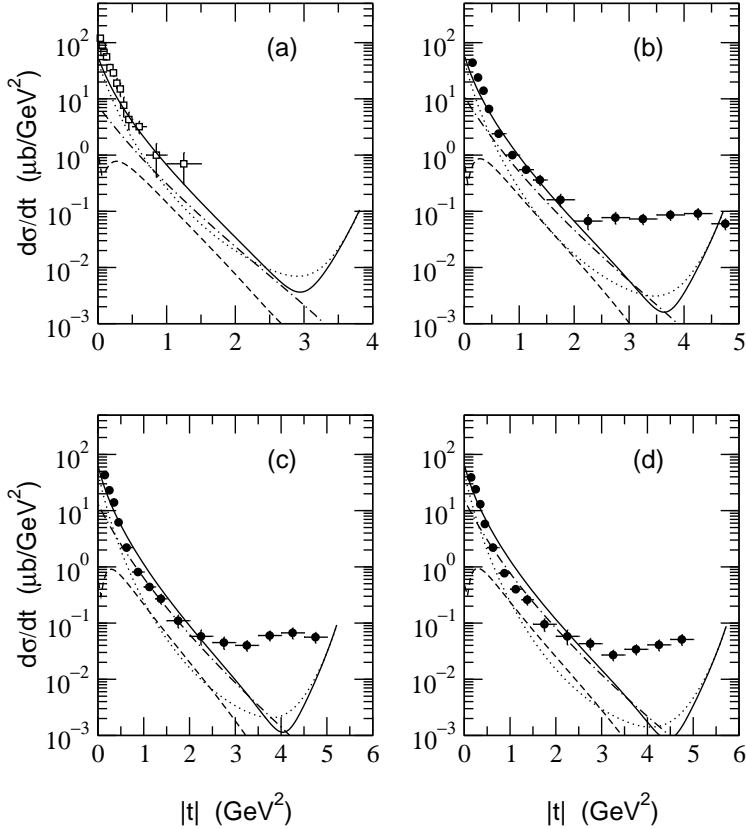}
\caption{Differential cross sections of model (B) at $E_\gamma = $ (a) 2.8,
(b) 3.28, (c) 3.55, and (d) 3.82 GeV. The dot-dashed lines are from $f_2$
exchange, the dashed lines are from $2\pi$ and $\sigma$ exchanges, and the
dotted lines are from the other processes, i.e., without $f_2$, $2\pi$, and
$\sigma$ exchanges.
The solid lines are the full calculation. Experimental data are from
Ref.~\cite{BCGG72} (open squares) and Ref.~\cite{CLAS01b} (filled circles).}
\label{fig:model-B}
\end{figure}

The differential cross sections calculated from model (B) are shown in
Fig.~\ref{fig:model-B}.
The solid curves are the best fits to the data we could obtain by choosing
the cutoff parameters given in Eq.~(\ref{eq:cut}) for the $f_2$ exchange.
In the same figures, we also show the contributions from the $f_2$
exchange (dot-dashed curves), $2\pi$ and $\sigma$ exchanges (dashed curves),
and the rest of the production mechanisms (dotted curves).
It is interesting to note that the $f_2$ exchange in model (B)
(dot-dashed curves in Fig.~\ref{fig:model-B}) drops faster than the
$\sigma$ exchange in model (A) (dot-ashed curves in Fig.~\ref{fig:model-A})
as $t$ increases.
On the other hand, the $2\pi$ and $\sigma$ exchanges (dashed curves in
Fig.~\ref{fig:model-B}) give a nontrivial contribution in large
$|t|$ region at lower energies but are suppressed as the energy
increases.
Therefore such effects are expected to be seen at energies very close to
the threshold.
As expected, the $\sigma$ meson exchange contribution is much suppressed
than in model (A).

\begin{figure}
\centering
\epsfig{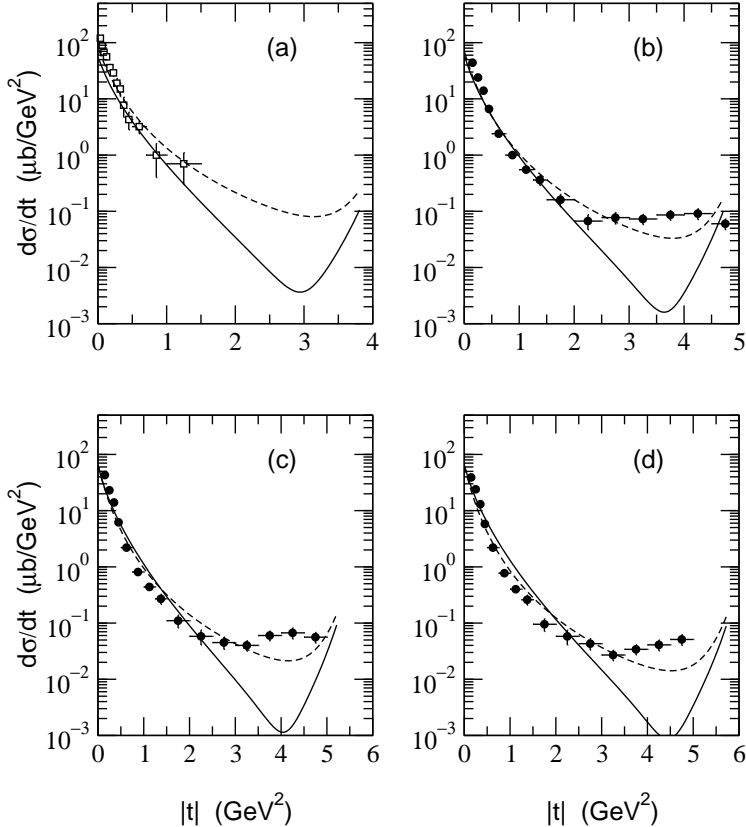}
\caption{Differential cross sections of model (A) and (B) at $E_\gamma = $
(1) 2.8, (b) 3.28, (c) 3.55, and (d) 3.82 GeV.
The dashed lines are the results of model (A) and the solid lines are
those of model (B).
Experimental data are from Refs.~\cite{CLAS01b,BCGG72}.}
\label{fig:dsdt}
\end{figure}

Thus we find that model (B) is comparable to model (A) that is the commonly
used $\sigma$ exchange model in fitting the differential cross section data
of SLAC and TJNAF.
In particular, the data at small $|t|$ ($ < 2 $ GeV$^2$) can be equally
well described by both models, as more clearly shown in Fig.~\ref{fig:dsdt},
where the full calculations of two models are compared.
On the other hand, both models cannot fit the data at large $|t|$ 
($> 2$ GeV$^2$).
But this is expected since we have not included $N^*$ and $\Delta^*$
excitation mechanisms which were found \cite{OTL01} to give significant
contributions to $\omega$ photoproduction at large $|t|$.
However, we will not address this rather non-trivial issue here.
The main difficulty here is that most of the resonance parameters associated
with isospin $T=3/2$ $\Delta^*$ resonances, which do not contribute to $\omega$
photoproduction, are not determined by Particle Data Group or
well-constrained by theoretical models.  
Before we use our model to determine a large number of resonance parameters
by fitting the existing limited data, it would be more desirable to further
test and improve the nonresonant amplitudes such as including more complete
calculations of $2\pi$-exchanges.
Hence, in this paper, we focus on exploring which experimental observables 
are useful for distinguishing more clearly the Model (B) from Model (A)
in the small $|t|$ ($ < 2 $ GeV$^2$) region where both models can
describe the differential cross section data to a large extent and
the $N^*$ and $\Delta^*$ effects are expected to be not important.
Experimental verifications of our prediction in this limited $t$ region
will be useful for understanding the non-resonant amplitudes
of $\rho$ photoproduction at low energies.

\begin{figure}
\centering
\epsfig{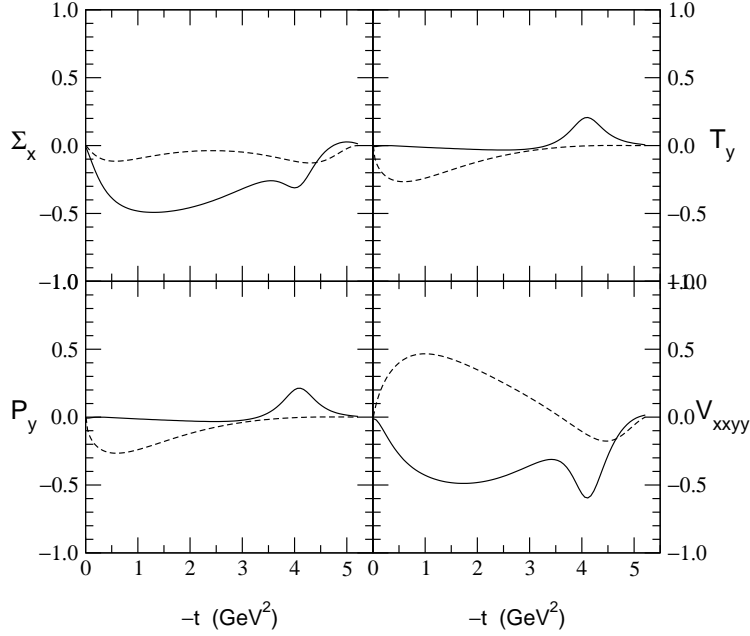}
\caption{Single spin asymmetries of model (A) and (B) at $E_\gamma = 
3.55$ GeV. Notations are the same as in Fig.~\ref{fig:dsdt}.
The definitions of the spin asymmetries are from Ref.~\cite{TOYM98}.}
\label{fig:3.55-pols}
\end{figure}

We have explored the consequences of the constructed models (A) and (B)
in predicting the spin asymmetries, which are defined, e.g., in 
Ref.~\cite{TOYM98}.
The results for the single spin asymmetries are shown in
Fig.~\ref{fig:3.55-pols} for $E_\gamma = 3.55$ GeV.
Clearly the single spin asymmetries including the target asymmetry
$T_y$, the recoiled proton asymmetry $P_y$, and the tensor asymmetry
$V_{xxyy}$ of the produced $\rho$ meson would be useful to distinguish the
two models and could be measured at the current experimental facilities.
Of course our predictions are valid mainly in the small $t$ region since
the $N^*$ and $\Delta^*$ excitations \cite{OTL01} or $G$-pole contributions
\cite{KV01a-KV02}, which are expected to be important at large $t$,
are not included in this calculation.

\begin{figure}
\centering
\epsfig{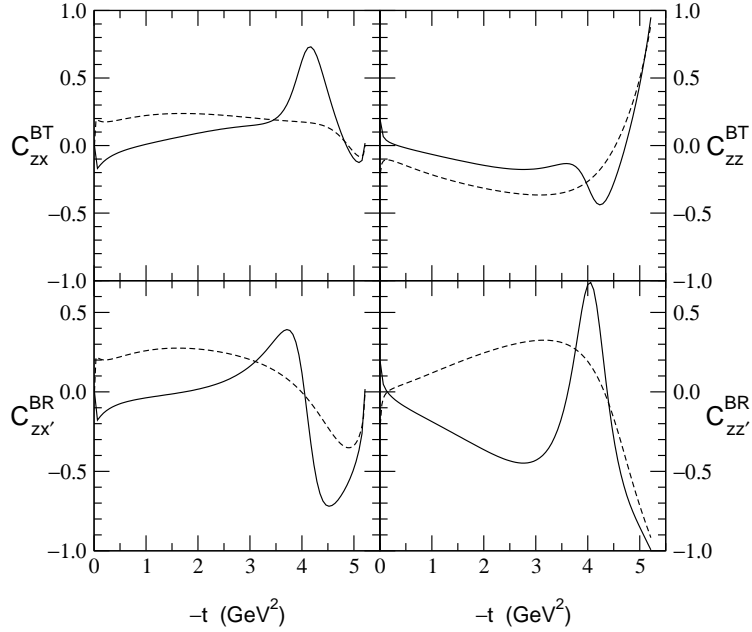}
\caption{Double spin asymmetries $C^{\rm BT}_{zx}$, $C^{\rm BT}_{zz}$,
$C^{\rm BR}_{zx'}$, and $C^{\rm BR}_{zz'}$ of model (A) and (B) at $E_\gamma = 
3.55$ GeV. Notations are the same as in Fig.~\ref{fig:dsdt}.}
\label{fig:3.55-pol-dble}
\end{figure}

Our predictions on the beam-target and beam-recoil double asymmetries
\cite{TOYM98} are given in Fig.~\ref{fig:3.55-pol-dble}.
Here again we can find significant differences between the two models
in the region of small $|t|$. 
Experimental tests of our predictions given in Figs.~\ref{fig:3.55-pols}
and \ref{fig:3.55-pol-dble}, therefore, will be useful in understanding
the non-resonant mechanisms of $\rho$ photoproduction.

\begin{figure}
\centering
\epsfig{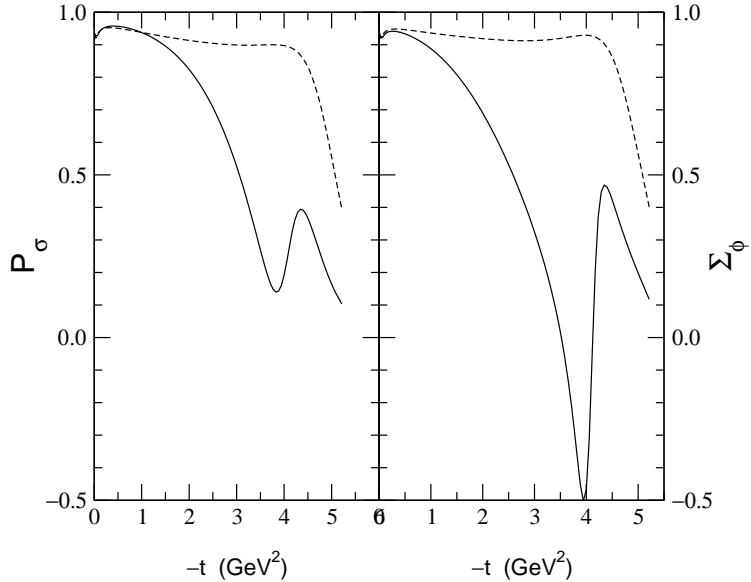}
\caption{Spin asymmetries $P_\sigma$ and $\Sigma_\phi$ of model (A) and (B)
at $E_\gamma = 3.55$ GeV. Notations are the same as in Fig.~\ref{fig:dsdt}.}
\label{fig:3.55-pol-decay}
\end{figure}

Since both the $\sigma$ and $f_2$ exchanges are natural parity
exchanges, it would be difficult to test them using parity asymmetry or
photon asymmetry that can be measured from the decay distribution of the
$\rho$ meson produced by polarized photon beam.
For completeness, we give the predictions of the two models on these
asymmetries in Fig.~\ref{fig:3.55-pol-decay}.
As expected, it is very hard to distinguish the two models in the
forward scattering angles with these asymmetries.

\section{Summary and Discussion}

In this paper we have re-examined the $\sigma$-exchange and
$f_2$-exchange mechanisms of $\rho$ photoproduction reactions.
It is found that the commonly employed $\sigma$-exchange amplitude is
weakened greatly if the $\sigma$ coupling constants are evaluated by
using the recent information about the $\rho \to \pi^0\pi^0\gamma$ decay
and the $\sigma NN$ coupling constant of Bonn potential.
This has led us to introduce the un-correlated $2\pi$ exchange amplitude
with $\pi N$ intermediate state.
This leading-order $2\pi$ exchange amplitude can be calculated realistically
using the coupling constants determined from the study of pion
photoproduction and the empirical width of $\rho \rightarrow \pi\pi$.

In the investigation of $f_2$-exchange mechanism, we evaluate its amplitude
using an effective Lagrangian which is constructed from the tensor
structure of the $f_2$ meson.
Phenomenological information together with tensor meson dominance and
vector meson dominance assumptions are used to estimate the $f_2$ 
coupling constants.
This approach, which is more consistent with the conventional
meson-exchange models, is rather different from the $f_2$-exchange model
of Laget \cite{Lage00}, where the $f_2$ interaction structure was borrowed
from that of Pomeron exchange assuming Pomeron-$f$ proportionality, i.e.,
$f_2$-photon analogy.

In comparing with the existing differential cross section data, we find
that a model with the constructed $2\pi$, $\sigma$, and $f_2$ exchanges is
comparable to the commonly used $\sigma$-exchange model in which the
$\sigma$ coupling parameters are simply adjusted to fit data.
Both models can describe the data equally well in the small $|t|$
($<2$ GeV$^2$) region, but fail at large $|t|$.
We suggest that experimental verifications of the predicted single and
double spin asymmetries in the small $|t|$ region will be useful for
distinguishing two models and improving our understanding of the
non-resonant amplitude of $\rho$ photoproduction.

Finally, we would like to emphasize that the present investigation is
just a very first step toward obtaining a complete dynamical exchange
model of $\rho$ photoproduction at low energies.
The next steps are to examine the additional $2\pi$-exchange mechanisms
due to, for example, $\omega N$ and $\pi \Delta$ intermediate states and
the crossed diagrams of Fig.~\ref{fig:2pi}.
The effects due to $N^*$ and $\Delta^*$ effects must be included for a
realistic understanding of the interplay between the non-resonant and
resonant amplitudes.
Theoretical predictions of the resonance parameters associated with
$\Delta^*$ resonance states will be highly desirable for making progress
in this direction.

\acknowledgments

Y.O. is grateful to the Physics Division of Argonne National Laboratory
for the hospitality during his stay.
The work of Y.O. was supported by Korea Research Foundation Grant
(KRF-2002-015-CP0074) and T.-S.H.L. was supported by
U.S. DOE Nuclear Physics Division Contract No. W-31-109-ENG-38.

\appendix*

\section{Tensor meson dominance and $\bm{f_2}$-hadron interactions}

The free Lagrangian and the propagator of the tensor meson were studied
in Refs. \cite{SW63,Wei64b,Chang66,BGS94,Toub96}.
The propagator of the tensor meson which has momentum $p$ reads
\begin{equation}
G^{\mu\nu;\rho\sigma} = \frac{1}{p^2 - M_f^2 + i\epsilon}
P^{\mu\nu;\rho\sigma},
\end{equation}
where $M_f$ is the tensor meson mass and
\begin{equation}
P^{\mu\nu;\rho\sigma} = \frac12 \left( \bar{g}^{\mu\rho}
\bar{g}^{\nu\sigma} + \bar{g}^{\mu\sigma} \bar{g}^{\nu\rho} \right) -
\frac13 \bar{g}^{\mu\nu} \bar{g}^{\rho\sigma},
\end{equation}
with
\begin{equation} 
\bar{g}_{\mu\nu} = -g_{\mu\nu} + \frac{p_\mu p_\nu}{M_f^2}.
\label{gbar}
\end{equation}

\subsection{$\bm{f_2\pi\pi}$ coupling}

The effective Lagrangian for $f_2 \pi\pi$ interaction reads
\cite{PSMM73}
\begin{equation}
\mathcal{L}_{f\pi\pi} = - \frac{2 G_{f\pi\pi}}{M_f} \partial_\mu
\bm{\pi} \cdot \partial_\nu \bm{\pi} f^{\mu\nu},
\label{fpipi}
\end{equation}
where $f^{\mu\nu}$ is the $f_2$ meson field.
This gives the $f_2\pi\pi$ vertex function as
\begin{equation}
V_{f\pi\pi} = -\frac{G_{f\pi\pi}}{M_f} (p_a + p_c)_\mu (p_a+p_c)_\nu
\epsilon^{\mu\nu}(\lambda_f),
\label{Vfpipi}
\end{equation}
where $p_a$ and $p_c$ are the incoming and outgoing pion momentum,
respectively. The minus sign in the Lagrangian (\ref{fpipi}) is to be
consistent with the tensor meson dominance \cite{Suzu93}.
The Lagrangian (\ref{fpipi}) gives the $f_2\to \pi\pi$ decay width as
\begin{equation}
\Gamma (f_2\to\pi\pi)
= \frac{G_{f\pi\pi}^2}{80\pi} M_f \left( 1 - 4 \frac{M_\pi^2}{M_f^2}
\right)^{5/2}.
\end{equation}
Using the experimental data, $\Gamma (f_2\to\pi\pi)_{\rm expt.}
\approx 156.9$ MeV \cite{PDG02}, we obtain
\begin{equation}
\frac{G_{f\pi\pi}^2}{4\pi} \approx 2.64,
\label{eq:Gfpp}
\end{equation}
which gives $G_{f\pi\pi} \approx 5.76$.

\subsection{Tensor meson dominance}

The tensor meson dominance (TMD) is an assumption of meson pole dominance for
matrix elements of the energy momentum tensor just as the vector meson
dominance (VMD) is a pole dominance of the electromagnetic current.
By using TMD, one can determine the universal coupling constant of the $f_2$
meson from its decay into two pions, which can then be used to determine
the $f_2 NN$ and $f_2 VV$ couplings.
When combined with VMD, this also allows us to estimate the
$f\gamma\gamma$ and $fV\gamma$ vertices.
It is interesting to note that the TMD underestimates the empirical
$f_2 NN$ coupling while it overestimates the $f_2\to\gamma\gamma$ decay width.
But it shows that the $f_2$ couplings with hadrons and photon can be
understood by TMD and VMD at least qualitatively.
Here, for completeness, we briefly review the method of
Refs.~\cite{Renn70,Renn71} to illustrate how to use TMD to get the
$f_2$-hadron couplings.

Let us first apply TMD to spinless particles \cite{Renn70,Raman71a}. The
energy-momentum tensor between spinless particles can be written as
\begin{equation}
\langle p | \theta^{\mu\nu}(0) | p' \rangle = F_1(\Delta^2) \Sigma_\mu
\Sigma_\nu + F_2(\Delta^2) (\Delta_\mu \Delta_\nu - g_{\mu\nu} \Delta^2
),
\label{spin0}
\end{equation}
with $\Sigma_\mu = (p+p')_\mu$ and $\Delta_\mu = (p-p')_\mu$.
Then with the covariant normalization one has
\begin{equation}
\langle p | \int \theta_{00}(x) d^3x | p \rangle = E N_p,
\end{equation}
where $N_p$ is the normalization constant.
By comparing with Eq.~(\ref{spin0}), one can find
\begin{equation}
F_1(0) = \frac12.
\end{equation}

Now we define the effective couplings for tensor mesons as
\begin{equation}
\langle f | \theta_{\mu\nu}(0) | 0 \rangle =
g_f^{} M_f^3 \epsilon_{\mu\nu}, \qquad
\langle p | f | p' \rangle = -\epsilon_{\mu\nu} \Sigma^\mu \Sigma^\nu
\frac{G_{fpp}}{M_f^{}},
\label{f0}
\end{equation}
where the latter equation is consistent with Eq.~(\ref{Vfpipi}).
The pole dominance gives
\begin{eqnarray}
\langle p | \theta_{\mu\nu} | p' \rangle &=& 
\sum_f \langle p | f | p' \rangle
\langle f | \theta_{\mu\nu} |0 \rangle \frac{1}{\Delta^2 - M_f^2}
\nonumber \\
&=& -\sum_fg_f^{} M_f^3 \epsilon_{\mu\nu} \epsilon^*_{\alpha\beta}
\Sigma^\alpha
\Sigma^\beta \frac{G_{fpp}}{M_f} \frac{1}{\Delta^2 - M_f^2}
\nonumber\\
&=& -\sum_f\frac{g_f^{} M_f^2 G_{fpp}}{\Delta^2 - M_f^2} \left( \Sigma_\mu
\Sigma_\nu - \frac13 g_{\mu\nu} \Sigma^2 + \frac13 \frac{\Delta_\mu
\Delta_\nu}{M_f^2} \Sigma^2 \right),
\end{eqnarray}
which leads to
\begin{equation}
F_1(\Delta^2) = -\sum_f \frac{g_f^{} M_f^2 G_{fpp}}{\Delta^2 - M_f^2}.
\end{equation}
Thus we have
\begin{equation}
F_1(0) = \sum_f g_f G_{fpp} = \frac12.
\label{eq:f10}
\end{equation}

It should be noted that the sum of Eq.~(\ref{eq:f10}) contains tensor
meson nonet, i.e., $f_2(1270)$ and $f_2'(1525)$.
But in the case of the $f_2\pi\pi$ coupling, if we assume the ideal mixing
between the $f_2(1270)$ and the $f_2'(1525)$, the $f_2'(1525)$
decouples by the OZI rule.
Therefore we obtain $G_{f'\pi\pi} \approx 0$, and the universal coupling
constant $g_f^{}$ is determined as
\begin{equation}
g_f^{} = \frac{1}{2 G_{f\pi\pi}^{}} \approx 0.087,
\label{gf}
\end{equation}
using the value of Eq.~(\ref{eq:Gfpp}).

With the universal coupling constant $g_f^{}$ determined above, one can now
use it to estimate the $f_2 NN$ coupling.
For this purpose, we apply TMD to spin-1/2 baryon state.
The energy-momentum tensor of the spin-1/2 baryons can be written as
\begin{eqnarray}
\langle p | \theta_{\mu\nu} (0) |p' \rangle &=& \bar{u}(p) \left\{ \frac14
(\gamma_\mu \Sigma_\nu + \gamma_\nu \Sigma_\mu ) F_1(\Delta^2)
\right. \nonumber \\ && \mbox{} \left. \qquad \mbox{} +
\frac{\Sigma_\mu \Sigma_\nu}{4M_N} F_2(\Delta^2)  + (\Delta_\mu
\Delta_\nu - g_{\mu\nu} \Delta^2 ) F_3(\Delta^2) \right\} u(p').
\end{eqnarray}
With the covariant normalization, the conditions
\begin{eqnarray}
&& \langle p | \int \theta_{00}(x) d^3x | p \rangle = E N_p, \nonumber \\
&& \langle p, {\bf p} = 0, s_3 = + \textstyle\frac12 |
\int\{ x_1 \theta_{02}(x) - x_2 \theta_{01}(x) \} d^3x | p ,
{\bf p} = 0, s_3 = + \frac12 \rangle = \frac12 N_p, \nonumber \\
\label{theta:con}
\end{eqnarray}
give
\begin{equation}
F_1(0) = 1, \qquad F_2(0) = 0.
\end{equation}

Now using the form for $f_2NN$ coupling in Eq.~(\ref{VfNN}), assuming the
pole dominance gives the following relations:
\begin{eqnarray}
-1 &=& 4 g_f^{} G_{fNN} \frac{M_f}{M_N} + 4 g_{f'}^{} G_{f'NN}
\frac{M_{f'}}{M_N}, \nonumber \\
0 &=& 4 g_f^{} F_{fNN} \frac{M_f}{M_N} + 4 g_{f'}^{} F_{f'NN}
\frac{M_{f'}}{M_N}.
\end{eqnarray}
Again by assuming the decoupling of the $f_2'$ from the nucleon coupling,
we can have \cite{Renn70}
\begin{eqnarray}
G_{fNN}^{} &=& \frac{1}{4g_f} \frac{M_p}{M_f} = \frac{G_{f\pi\pi}}{2} 
\frac{M_p}{M_f} \approx 2.12, \nonumber \\
F_{fNN}^{} &=& 0.
\end{eqnarray}
This gives $G_{fNN}^{2}/4\pi \approx 0.38$ as shown in Table \ref{tab:fnn},
which is smaller than the values estimated by $\pi N$ dispersion
relations by an order of magnitude.
It should also be noted that the values estimated by $\pi N$ dispersion
relations may be affected by the inclusion of other meson exchanges.
More rigorous study in this direction is, therefore, highly desirable.

\subsection{$\bm{f_2VV}$ coupling}

Before we discuss $f_2\gamma\gamma$ and $f_2V\gamma$ couplings, we first
apply TMD to $f_2VV$ coupling, where $V$ stands for vector mesons.
The energy-momentum tensor between identical vector mesons contains six
independent matrix elements \cite{Renn71},
\begin{eqnarray}
\langle V | \theta_{\mu\nu} | V' \rangle &=& \mathcal{G}_1(\Delta^2)
(\epsilon \cdot \epsilon') \Sigma_\mu \Sigma_\nu
+ \mathcal{G}_2(\Delta^2) (\epsilon \cdot \Sigma) (\epsilon' \cdot \Sigma)
\Sigma_\mu \Sigma_\nu
\nonumber \\ && \mbox{}
+ \mathcal{G}_3(\Delta^2) \left\{ (\epsilon \cdot \Sigma)
\epsilon'_\mu \Sigma_\nu
+ (\epsilon \cdot \Sigma) \epsilon'_\nu \Sigma_\mu + (\epsilon' \cdot
\Sigma) \epsilon_\mu \Sigma_\nu + (\epsilon' \cdot \Sigma) \epsilon_\nu
\Sigma_\mu \right\}
\nonumber \\ && \mbox{}
+ \mathcal{G}_4(\Delta^2) \left\{ (\epsilon \cdot \Delta)
\epsilon_\mu' \Delta_\nu
+ (\epsilon \cdot \Delta) \epsilon_\nu' \Delta_\mu + (\epsilon' \cdot
\Delta ) \epsilon_\mu \Delta_\nu + (\epsilon' \cdot \Delta) \epsilon_\nu
\Delta_\mu
\right. \nonumber \\ && \mbox{} \qquad \qquad \left.
- 2(\epsilon \cdot \Delta)(\epsilon' \cdot \Delta) g_{\mu\nu}
- \Delta^2 (\epsilon_\mu \epsilon_\nu' + \epsilon_\mu' \epsilon_\nu)
\right\}
\nonumber \\ && \mbox{}
+ \mathcal{G}_5(\Delta^2) (\epsilon \cdot \epsilon')
( \Delta_\mu \Delta_\nu - \Delta^2 g_{\mu\nu})
\nonumber \\ && \mbox{}
+ \mathcal{G}_6(\Delta^2) (\epsilon \cdot \Sigma) (\epsilon' \cdot \Sigma)
( \Delta_\mu \Delta_\nu - g_{\mu\nu} \Delta^2 ),
\label{theta:TVV}
\end{eqnarray}
where $\Sigma_\mu = (p+p')_\mu$, $\Delta_\mu = (p-p')_\mu$ and
$\epsilon, \epsilon'$ are the polarization vectors of $V$ and $V'$,
respectively.
Then the conditions like Eq.~(\ref{theta:con}) give
\begin{equation}
\mathcal{G}_1(0) = - \frac12, \qquad \mathcal{G}_3(0) = \frac12.
\label{fVV:con}
\end{equation}

In the pole model, the form factors $\mathcal{G}_1(\Delta^2) \dots
\mathcal{G}_4(\Delta^2)$ are dominated by tensor meson poles.
Because of the symmetry property of the tensor meson, we have generally
four $f_2VV$ coupling vertices:
\begin{eqnarray}
\langle V | f | V' \rangle &=& \frac{G_1}{M_f} (\epsilon
\cdot \epsilon') (\Sigma_\mu \Sigma_\nu f^{\mu\nu})
+ \frac{G_2}{M_f^3} (\epsilon \cdot \Sigma) (\epsilon' \cdot \Sigma)
(\Sigma_\mu \Sigma_\nu f^{\mu\nu})
\nonumber \\ && \mbox{}
+ \frac{G_3}{M_f} \left\{ (\epsilon \cdot \Sigma) \epsilon'_\mu \Sigma_\nu
+ (\epsilon \cdot \Sigma) \epsilon'_\nu \Sigma_\mu + (\epsilon' \cdot
\Sigma) \epsilon_\mu \Sigma_\nu + (\epsilon' \cdot \Sigma) \epsilon_\nu
\Sigma_\mu \right\} f^{\mu\nu}
\nonumber \\ && \mbox{}
+ \frac{G_4}{M_f}  
(- \Delta^2) (\epsilon_\mu \epsilon_\nu' + \epsilon_\mu' \epsilon_\nu)
f^{\mu\nu},
\label{eq:fVV}
\end{eqnarray}
while we have used $\Delta^\mu f_{\mu\nu} = f^\mu_\mu = 0$ in writing
the $G_4$ term.
For our later use, an effective vertex $H(\Delta^2,p^2,p'^2)$ is
introduced to replace $-\frac{G_4}{M_f} \Delta^2$ as \cite{Renn71}
\begin{equation}
H(\Delta^2,p^2,p'^2) = \frac{G_4}{M_f} \left\{ - \Delta^2 + \alpha (p^2
+ p'^2 - 2M_V^2) \right\}.
\label{H}
\end{equation}

Now we use the pole dominance again using Eq.~(\ref{f0}) to find
\begin{equation}
\mathcal{G}_1(\Delta^2) = \frac{g_f M_f^2 G_1}{\Delta^2 - M_f^2},
\end{equation}
which leads to
\begin{equation}
\frac12 = g_f G_1, \qquad
G_3 = - G_1,
\end{equation}
combined with Eq.~(\ref{fVV:con}).
Therefore, with Eq.~(\ref{gf}) we get
\begin{equation}
G_1 = - G_3 = G_{f\pi\pi} \approx 5.76.
\end{equation}
The above relation should hold for $f_2\rho\rho$ and $f_2\omega\omega$. The
SU(3) symmetry and the ideal mixing give
\begin{equation}
G_1(f_2'\phi\phi) = \sqrt2 G_1(f_2\rho\rho), \qquad G_1(f_2'\phi\omega) =
G_1(f_2\phi\phi) = 0.
\end{equation}

Note that two couplings $G_1$ and $G_3$ are determined by TMD but $G_2$
and $G_4$ cannot be estimated without further assumptions.

\subsection{$\bm{f_2\gamma\gamma}$ and $\bm{f_2V\gamma}$ couplings}

\begin{figure}
\centering
\epsfig{file=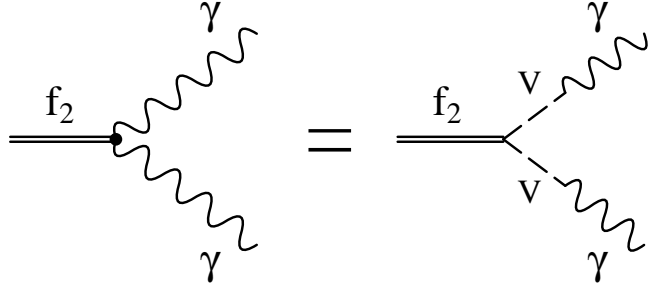, width=0.5\hsize}
\caption{$f_2 \to \gamma\gamma$ decay in vector meson dominance.}
\label{fig:fggdecay}
\end{figure}

The remaining two couplings $G_2$ and $G_4$ of Eq.~(\ref{eq:fVV}) are
estimated by using VMD and gauge invariance.
We consider $f_2 \to \gamma \gamma$ using VMD as illustrated in
Fig.~\ref{fig:fggdecay}.

By using $\epsilon \cdot k = \epsilon' \cdot k' = 0$ and VMD, we have
\begin{eqnarray}
\langle \gamma(k) \gamma(k') | f \rangle &=&
\frac{e^2}{(k^2 - M_V^2)(k'^2-M_V^2)} \left\{
\frac{\tilde{G}_1}{M_f} (\epsilon
\cdot \epsilon') (k-k')_\mu (k-k')_\nu f^{\mu\nu}
\right.\nonumber \\ && \mbox{}
- \frac{\tilde{G}_2}{M_f^3} (\epsilon \cdot k') (\epsilon' \cdot k)
(k-k')_\mu (k-k')_\nu f^{\mu\nu}
\nonumber \\ && \mbox{}
+ \frac{\tilde{G}_3}{M_f}
\left[ -(\epsilon \cdot k') \epsilon'_\mu (k-k')_\nu
- (\epsilon \cdot k') \epsilon'_\nu (k-k')_\mu
\nonumber \right. \\ && \left.\mbox{} \qquad\qquad
+ (\epsilon' \cdot
k) \epsilon_\mu (k-k')_\nu + (\epsilon' \cdot k) \epsilon_\nu
(k-k')_\mu \right] f^{\mu\nu}
\nonumber \\ && \mbox{} \left.
+ \frac{\tilde{G}_4}{M_f}
[- M_f^2 + \alpha(k^2 + k'^2 - 2M_V^2)]
(\epsilon_\mu \epsilon_\nu' + \epsilon_\mu' \epsilon_\nu)
f^{\mu\nu} \right\},
\label{fgg}
\end{eqnarray}
where we have introduced the notation
\begin{equation}
\tilde{G}_i = \left(\frac{M_\rho^2}{f_\rho}\right)^2 G_i^{f\rho\rho}
+ \left(\frac{M_\omega^2}{f_\omega}\right)^2 G_i^{f\omega\omega},
\label{eq:Gtilde}
\end{equation}
with
\begin{equation}
\langle 0 | j_\mu^{\rm em} | V \rangle = \frac{M_V^2}{f_V}
\epsilon_\mu(V).
\end{equation}

Because of isospin, there is no mixing between the intermediate $\rho$
and $\omega$ mesons.
By looking at the amplitude (\ref{fgg}), however, one can find that it is
not gauge invariant, i.e., it does not vanish when replacing $\epsilon_\mu$
by $k_\mu$.
This gives a constraint on the couplings.
The most general form for $f_2\gamma\gamma$ satisfying gauge invariance has
two independent couplings as \cite{Renn71}
\begin{eqnarray}
\langle \gamma(k) \gamma(k') | f \rangle &=& \frac{e^2}{M_V^4} \Bigl\{
A \left[ (\epsilon\cdot\epsilon') (k \cdot k') - (\epsilon \cdot k')
(\epsilon' \cdot k) \right] (k-k')_\mu (k-k')_\nu f^{\mu\nu}
\nonumber \\ && \mbox{}
+ B [ ( \epsilon \cdot \epsilon') (k-k')_\mu (k-k')_\nu + \epsilon_\mu'
(k-k')_\nu (\epsilon \cdot k') + \epsilon_\nu' (k-k')_\mu (\epsilon
\cdot k')
\nonumber \\ && \mbox{} \qquad
- \epsilon_\mu (k-k')_\nu (\epsilon' \cdot k) - \epsilon_\nu (k-k')_\mu
(\epsilon' \cdot k)
\nonumber \\ && \mbox{} \qquad
- 2 (k \cdot k') (\epsilon_\mu \epsilon'_\nu + \epsilon'_\mu
\epsilon_\nu) ] f^{\mu\nu} \Bigr\},
\end{eqnarray}
which then gives
\begin{eqnarray}
\frac{\tilde{G}_1}{M_f} &=& (k \cdot k') A + B, \nonumber \\
\frac{\tilde{G}_2}{M_f^3} &=& A , \nonumber \\ 
\frac{\tilde{G}_3}{M_f} &=& -B , \nonumber \\
\frac{\tilde{G}_4}{M_f} \left[ -M_f^2 + \alpha (k^2 + k'^2 - 2M_V^2)
\right] &=& - 2 (k \cdot k') B = (k^2 + k'^2 - M_f^2 ) B.
\end{eqnarray}
Solving this system at $k^2 = k'^2 = 0$ and $\tilde{G}_1 = -
\tilde{G}_3$ gives
\begin{equation}
A = \tilde{G}_2 = 0.
\end{equation}
Since gauge invariance applies to isoscalar and isovector photons
separately, we get $G_2 = 0$ for $V = \rho, \omega$.
Still we do not fix $\tilde{G}_4$ and $\alpha$, but have a constraint,
\begin{equation}
\tilde{G}_4 (M_f^2 + 2 \alpha M_V^2) = \tilde{G}_1 M_f^2.
\label{eq1}
\end{equation}

\begin{figure}
\centering
\epsfig{file=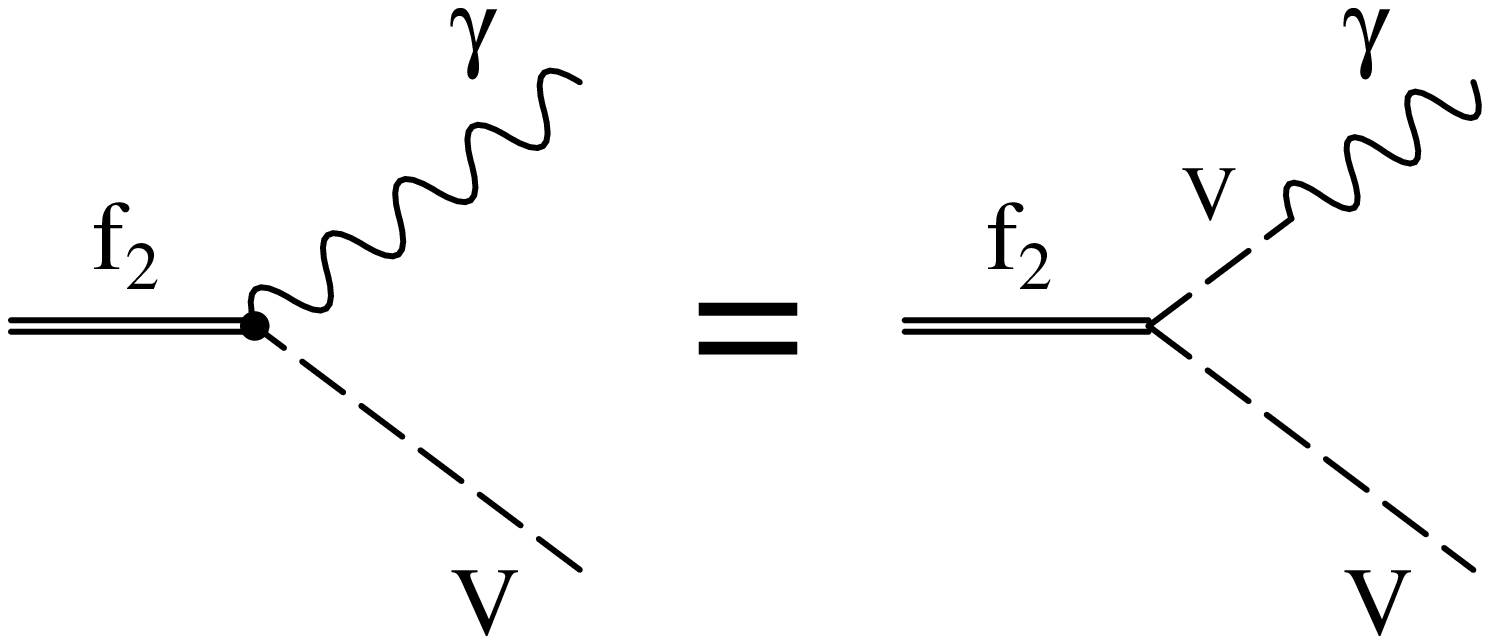, width=0.5\hsize} 
\caption{$f_2 \to V\gamma$ decay in vector meson dominance.}
\label{fig:fvgdecay}
\end{figure}

To complete the model, let us finally consider $fV\gamma$ vertex.
Here again, we use the VMD as in Fig.~\ref{fig:fvgdecay}. The gauge
invariance of the vertex at $k^2 = M_V^2$ and $k'^2 = 0$ leads to
\begin{equation}
\tilde{G}_4 (M_f^2 + \alpha M_V^2) = \tilde{G}_1 (M_f^2-M_V^2).
\label{eq2}
\end{equation}
Then solving the coupled equations (\ref{eq1}) and (\ref{eq2}) gives
\cite{Renn71}
\begin{equation}
\tilde{G}_4 = \tilde{G}_1 \frac{M_f^2 - 2M_V^2}{M_f^2}, \qquad
\alpha = \frac{M_f^2}{M_f^2 - 2M_V^2}.
\end{equation}
Thus we have determined all couplings of Eq.~(\ref{eq:fVV}) with the
relation (\ref{eq:Gtilde}).

The above procedure shows that the $f_2\gamma\gamma$ and $f_2V\gamma$
vertices can be written with two form factors because of gauge
invariance, which read
\begin{eqnarray}
\langle \gamma(k) \gamma(k') | f_2 \rangle &=& \frac{1}{M_f}
\epsilon^\kappa \epsilon'^\lambda f^{\mu\nu} A_{\kappa \lambda
\mu\nu}^{f\gamma\gamma}(k,k'), \nonumber \\
\langle \gamma(k) V(k') | f_2 \rangle &=& \frac{1}{M_f}
\epsilon^\kappa \epsilon'^\lambda f^{\mu\nu} A_{\kappa \lambda
\mu\nu}^{fV\gamma}(k,k'),
\label{FVV}
\end{eqnarray}
where
\begin{eqnarray}
A^{f\gamma\gamma}_{\kappa\lambda\mu\nu} (k,k') &=&
\frac{f^{}_{f\gamma\gamma}}{M_f^3} \left[ g_{\kappa \lambda} (k\cdot k')
- k'_\kappa k_\lambda \right] (k-k')_\mu (k-k')_\nu
\nonumber \\ && \mbox{}
+ g_{f\gamma\gamma}^{} [ g_{\kappa\lambda} (k-k')_\mu (k-k')_\nu
+ g_{\lambda\mu} k'_\kappa (k-k')_\nu
+ g_{\lambda\nu} k'_\kappa (k-k')_\mu
\nonumber \\ && \mbox{} \qquad\qquad
- g_{\kappa\mu} k_\lambda (k-k')_\nu
- g_{\kappa\nu} k_\lambda (k-k')_\mu
\nonumber \\ && \mbox{} \qquad\qquad
-2 k \cdot k' ( g_{\kappa\mu} g_{\lambda\nu} + g_{\kappa\nu}
g_{\lambda\mu}) ].
\label{AFVV}
\end{eqnarray}
The vertex function $A_{\kappa \lambda \mu\nu}^{fV\gamma}(k,k')$ can be
obtained from $A^{f\gamma\gamma}_{\kappa\lambda\mu\nu}$ by replacing
$f^{}_{f\gamma\gamma}$ and $g_{f\gamma\gamma}^{}$ by
$g_{fV\gamma}^{}$ and $g_{fV\gamma}^{}$, respectively.

With Eqs. (\ref{FVV}) and (\ref{AFVV}), we can obtain the $f_2 \to
\gamma\gamma$ decay width as%
\footnote{Here we do not agree with the decay width formula of Ref.
\cite{Tera90}.}
\begin{equation}
\Gamma(f_2 \to \gamma\gamma) = \frac{M_f}{20\pi} \left( \frac{1}{24}
f_{f\gamma\gamma}^2 + g_{f\gamma\gamma}^2 \right).
\end{equation}
Then TMD and VMD give \cite{Renn71}
\begin{equation}
f^{}_{f\gamma\gamma} = 0, \qquad
g^{}_{f\gamma\gamma} = e^2 \left( \frac{1}{f_\rho^2} +
\frac{1}{f_\omega^2} \right) G_{fVV}.
\end{equation}
The vector meson decay constants are $f_\rho = 5.33$, $f_\omega = 15.2$,
and $f_\phi = 13.4$.
By noting that TMD gives $G_{fVV} = G_{f\pi\pi}$, we get
\begin{equation}
\Gamma(f_2 \to \gamma\gamma) \approx 8.8 \mbox{ keV},
\end{equation}
while its experimental value is $\Gamma(f_2\to \gamma\gamma)_{\rm expt.} =
2.6 \pm 0.24$ keV.
Thus we can find that this procedure overestimates the experimental value
by a factor of $3 \sim 4$.

The decay width of $f_2 \to V\gamma$ can be computed using
Eqs.~(\ref{FVV}) and (\ref{AFVV}) as
\begin{eqnarray}
\Gamma(f_2 \to V\gamma) &=& \frac{M_f}{10\pi} (1-x)^3 \left\{
\frac{1}{24} |f^{}_{fV\gamma}|^2 (1-x)^4 - (f^{}_{fV\gamma}
g^*_{fV\gamma} + f^*_{fV\gamma} g^{}_{fV\gamma}) \frac{x(1-x)^2}{12}
\right. \nonumber \\ && \mbox{} \left.
+ |g^{}_{fV\gamma}|^2 \left( 1 + \frac{x}{2} + \frac{x^2}{6} \right)
\right\},
\end{eqnarray}
where $x = M_V^2/M_f^2$.
TMD combined with VMD gives \cite{Renn71}
\begin{equation}
f^{}_{fV\gamma} = 0, \qquad
g^{}_{fV\gamma} = \frac{e}{f_V} G_{fVV}.
\end{equation}
This leads to
\begin{equation}
\Gamma(f_2 \to \rho \gamma) / \Gamma(f_2 \to \omega \gamma) =
\frac{g_{f\rho\gamma}^2}{g_{f\omega\gamma}^2} =
\frac{f_\omega^2}{f_\rho^2} = 8.14 \pm 1.2
\end{equation}
and \cite{Suzu93}
\begin{equation}
\Gamma(f_2 \to \rho \gamma) / \Gamma(f_2 \to \gamma \gamma) = 2
\frac{g_{f\rho\gamma}^2}{g_{f\gamma\gamma}^2} (1-x)^3 \left( 1 +
\frac{x}{2} + \frac{x^2}{6} \right) = 155.
\end{equation}
Those quantities are not measured yet.
Therefore, measuring those quantities will be very useful to test TMD
and VMD.

\end{document}